\documentclass{osa-article}
\journal{oe}
\articletype{Research Article}
\usepackage{booktabs, makecell, siunitx, eqparbox}
\usepackage{textcomp}

\begin{document}

\title{Noise-robust single-pixel imaging in photon counting regime with a pulsed source}

\author{Junghyun Kim, \authormark{1} Sangkyung Lee, \authormark{1} Yonggi Jo,\authormark{1} Su-Yong Lee,\authormark{1} Taek Jeong,\authormark{1} Dongkyu Kim,\authormark{1} Duk Y. Kim,\authormark{1} Zaeill Kim,\authormark{1} and Yong Sup Ihn\authormark{1,*}}

\address{\authormark{1}, Agency for Defense Development, Daejeon 34186, Korea}
\email{\authormark{*}yong0862@add.re.kr}

\begin{abstract} 
We present a method to classically enhance noise-robustness of single-pixel imaging in photon counting regime with a pulsed source.
By using time-domain cross-correlations between temporal profiles of a pulsed source and received signals, our scheme classically imitates the noise rejection concept of quantum imaging.
Under a strong noise environment in which the background noise intensity is up to 120 times higher than the signal one, we compare three different images obtained by conventional, quantum-enhanced, and classically enhanced schemes.
The results show that the classically enhanced scheme can be remarkably robust against noise in image formation, which is comparable to the quantum scheme.
\end{abstract}

\section{Introduction}
Quantum correlations between photon-pairs generated via spontaneous parametric down-conversion (SPDC) have been widely used for quantum imaging to enhance the performance of the image formation over the classical limit \cite{NatPho10Brida, APL15Xu}.   
After it was demonstrated that the quantum advantages can still be valid in a lossy and noisy environment that degrades the initial quantum states \cite{Science08Lloyd, PRL08Tan, NP10Kacprowicz, PRL13Lopaeva, PRL15Zhang, PRA21Lee}, quantum correlations have been employed in imaging to achieve noise-robustness.
Recently, notable imaging methods that exploit a spatially correlated twin-beam and multi-pixel camera have been introduced and have demonstrated the noise-robustness of quantum illuminated imaging \cite{SciAdv19Defienne, SciAdv20Gregory}. 
Similarly, raster-scanned imaging and single-pixel imaging schemes have shown enhanced image qualities in a noisy environment by using temporal correlations between photon-pairs \cite{PRA08Shapiro, NP19Edgar,PRA19England, APL20Yang, 21Kim}.
The joint measurements of correlated photon-pairs filter out background noise photons outside the narrow coincidence window, significantly suppressing the noise contribution to the image reconstruction. 
As a result, these quantum approaches have led to substantial improvements in image formations against strong background noise that severely degrades the image qualities in conventional imaging.

Apart from quantum approaches, several classical methods to reduce the noise effect on photon-counting regime imaging have been introduced \cite{SciRep18Liu, AppOpt15Yang, OE20Liu}. 
By using a signal processing algorithm and time-correlated single-photon counting techniques, they successfully eliminated the effect of background noise, which was few times larger than the signal photon number, and enhanced image qualities over the conventional imaging method. Although demonstrating impressive classical methods for noise suppression in photon-counting regime, they did not show the performance enhancement against much stronger background noise, e.g. 100 times brighter than the signal. Furthermore, a comparison of quantum imaging and classically enhanced imaging under various noise intensities has not yet been reported.

In this work, we present a classical method to significantly enhance the noise-robustness of single-pixel imaging in photon counting regime by using a pulsed source. The basic concept of our pulsed source single-pixel imaging (PSPI) is to classically imitate the twin-beam-based noise suppression of quantum imaging. While quantum single-pixel imaging (QSPI) makes use of temporal correlations between signal and idler photons \cite{APL20Yang,21Kim}, PSPI utilizes time-domain cross-correlations between temporal profiles of photon source-driving pulses and received signals so that the noise photons outside the pulse width can be filtered out. Under a heavy noise environment, where the target object is subject to 120 times stronger background noise than the signal, PSPI demonstrates a remarkable improvement in noise rejection compared to conventional single-pixel imaging (CSPI). We also report a comparison of imaging performances of CSPI, QSPI, and PSPI in terms of signal-to-noise ratio (SNR) under different pulse widths and noise intensities. As a result, it is shown that a proper selection of pulse width allows PSPI to provide a comparable noise rejection to QSPI.

\section{Correlation-based noise-suppression in single-pixel imaging}

\begin{figure*}[b]
\centering
\includegraphics[width = 0.8\textwidth]{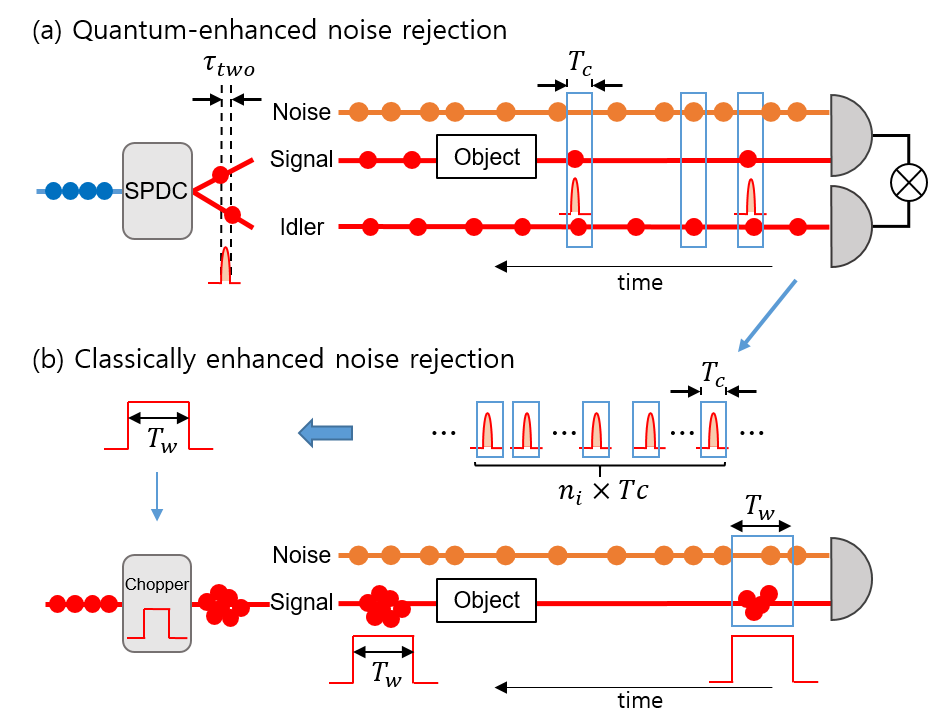}
\caption{Noise rejection concept of (a) quantum-enhanced imaging and (b) classically enhanced imaging. Quantum imaging distinguishes signal photons from noise photons by measuring coincidence events with correlated idler photons. Classically enhanced imaging makes use of a single pulse of signal photons and detect noise photons arriving within the pulse width.}\label{concept}
\end{figure*}

The classical imitation of quantum-enhanced noise rejection is visualized in Fig. \ref{concept}. In the quantum approach \cite{PRA19England, SciAdv20Gregory, 21Kim}, correlated photon-pairs with extremely short time separation $\tau_{two} (< ps)$ are generated via SPDC, and signal photons exist probabilistically in the time-domain. 
Joint measurements of signal and idler photons within a narrow coincidence window $T_c (> \tau_{two})$ distinguish signal photons from background noise photons. Only noise photons that accidentally arrive in the coincidence window contribute to the image reconstruction. Therefore, the total window allowed for noise photons can be given by the product of idler photon number $n_i$ and the coincidence window $T_c$. 

In our classical scheme, we use a single pulsed beam with pulse width $T_w$ instead of a probabilistically distributed photon stream in the time-domain.
Similar with quantum approach, we count only accidental noise photons arriving within the pulse width. 
Given one pulse per acquisition time, the total window for noise photons is the same as the pulse width $T_w$. 
Therefore, quantum-enhanced noise rejection based on temporal correlations of photon-pairs can be imitated by setting $T_w$ close to $n_iT_c$.

Before applying these noise suppression methods into imaging, we first formulate the single-pixel imaging process. Single-pixel imaging (SPI), or computational ghost imaging, obtains the target object's image by using spatially modulated light and intensity differences of modulated patterns \cite{PRA08Shapiro,NP19Edgar,OE20Liu}. Assume that we have a set of $N$ complete 2D basis patterns $\{P^{(1)}, P^{(2)},..., P^{(N)}\}$. If we illuminate the target object with pattern-modulated light, the signal intensity recieved from the target object depends on the pattern and an object shape. With a set of intensities $\{I^{(1)}, I^{(2)},..., I^{(N)}\}$ of which each component corresponds to one pattern, the image of the target object can be reconstructed by calculating the covariance $G^{(2)}$ between the modulation patterns and intensities:     
\begin{equation}\label{Cov}
G^{(2)}(i,j)\: = \:\langle P^{(k)}(i,j)\: I^{(k)}\rangle - \langle P^{(k)}(i,j)\rangle \langle I^{(k)}\rangle,
\end{equation}
where $\langle...\rangle$ is an average over $N$, and $i$ and $j$ are 2D pixel indices. 

\begin{figure*}[b]
\centering
\includegraphics[width = 0.5\textwidth]{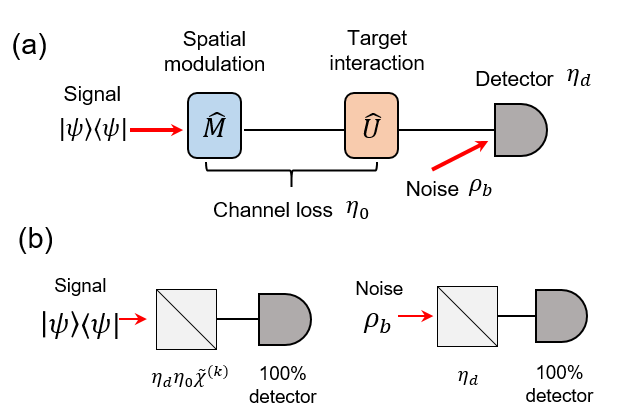}
\caption{(a) Schematic diagram of the SPI process in a noisy and lossy environment. Spatially modulated signal photons are sent to a target object with channel loss $\eta_0$. Then, the transmitted signal photons and incoherent thermal noise are measured by a detector with an efficiency $\eta_d$. (b) SPI process modeled by a BS. Signal photons enter an ideal detector after a BS with effective transmittance $\eta_d\eta_0\tilde{\chi}^{(k)}$, and noise photons impinge on the ideal detector after BS with transmittance $\eta_d$.}\label{BSmodel}
\end{figure*}

A general SPI process is shown in Fig. \ref{BSmodel}(a). 
$\bar{n}_s$ signal photons are sent to a spatial light modulator (SLM), which transforms the signal photon number into $\frac{1}{M}\sum_{i,j}^{M}P^{(k)}(i,j)\bar{n}_s$, where $M$ is the total number of pixels of the pattern.
After interaction with the target object, the photon number changes to $\frac{1}{M}\sum_{i,j}^{M}P^{(k)}(i,j)\chi(i,j)\bar{n}_s$, where $\chi(i,j)$ denotes the target object's spatial profile in 2D matrix.
Here, $\chi(i,j)=1$ if the target object exists at the $(i,j)_{th}$ pixel, and 0 if it does not.
In the presence of background noise photons $\bar{n}_b$, the detected photon number is given by $I^{(k)}$ = $\eta_0\eta_{d}\frac{1}{M}\sum_{i,j}P^{(k)}(i,j)\chi(i,j)\bar{n}_{s}$ + $\bar{n}_b\eta_d$ $=\eta_0\eta_{d}\tilde{\chi}^{(k)}\bar{n}_{s}$ + $\eta_d\bar{n}_b$ for the $k_{th}$ pattern. $\eta_0$ and $\eta_{d}$ refer to an overall channel loss and detector efficiency, respectively, and $\tilde{\chi}^{(k)}$ represents an overlapped portion between the target object and pattern.
Substituting $I^{(k)}$ into Eq. \eqref{Cov} and using Hadamard patterns as modulation basis, we obtain
\begin{equation}\label{recon}
\begin{split}
G^{(2)}(i,j)&= \eta_0\eta_d
\bar{n}_s/4M \:\:\:\: \text{if} \:\:\:\: \chi(i,j) = 1, \\
& = 0 \:\:\:\:\:\:\:\:\:\:\:\:\:\:\:\:\:\:\: \text{if}\:\:\:\: \chi(i,j) = 0,
\end{split}
\end{equation}
where the constant noise is removed from the covariance, and the object shape is perfectly reconstructed. 

Now let us consider the noise suppression together with photon statistics and photoelectric process of detectors. 
Using a practical detector model \cite{Optica17Rafsanjani}, the total SPI process can be simplified to an effective beam splitter (BS) and a 100\% efficient detector as shown in Fig. \ref{BSmodel}(b).
For signal photons, the effective BS transmittance is written as $\eta_d\eta_0\tilde{\chi}^{(k)}$.
For thermal background noise $\rho_b = \sum_{n=0}^{\infty} \frac{(\bar{n}_b)^n}{(\bar{n}_b+1)^{n+1}} |n\rangle\langle n|$, the density matrix transforms into $\sum_{n=0}^{\infty} \frac{(\bar{n}_b\eta_d)^n}{(\bar{n}_b\eta_d+1)^{n+1}} |n\rangle\langle n|$ after interacting with the BS.
This density operator implies that the photon statistics of thermal noise arriving at the detector represents the Bose-Einstein distribution with mean $\bar{N}_b =$ $\bar{n}_b\eta_d$ and variance $\sigma_b^2 = $ $\bar{n}_b\eta_d(\bar{n}_b\eta_d+1)$.   

In QSPI, the down-converted photon-pair is a two-mode squeezed vacuum state $|\text{TMSV}\rangle$\cite{PRL08Tan}. 
The signal mode is sent to the target object and the idler mode is directly sent to the detector to herald signal photons. Tracing out the idler mode of $|\text{TMSV}\rangle$, signal photons follow the Bose-Einstein distribution with mean $\bar{N}_{sq} = $$\eta_d\eta_0\tilde{\chi}^{(k)}\bar{n}_{sq}$ and variance $\sigma_{sq}^2 = $$\eta_d\eta_0\tilde{\chi}^{(k)}\bar{n}_{sq}(\eta_d\eta_0\tilde{\chi}^{(k)}\bar{n}_{sq}+1)$. We put the additional subscript $q$ to indicate the quantum light.
Likewise, idler photons represent the same distribution with mean $\bar{N}_i =$ $\eta_d\bar{n}_{sq}$ and variance $\sigma_i^2 = $ $\eta_d\bar{n}_{sq}(\eta_d\bar{n}_{sq}+1)$.
Note that the photon statistics is valid within the coherence time. 
The accumulated photon number $\mu_j$ during the acquisition time $T_{aq}$, which is much longer than the coherence time, can be calculated with the central limit theorem: 
\begin{equation}\label{numb}
\begin{split}
&\mu_{j} \sim N(L_{j}\bar{N}_{j}, \:\:L_j\sigma_{j}^2),\\
\end{split}
\end{equation}
where $j\in\{sq,i,b\}$, and $N(x,y)$ denotes the normal distribution with mean $x$ and variance $y$.
Here, the coherence times of signal photons, idler photons, and noise photons are $\tau_{sq}$, $\tau_{i}$, and $\tau_{b}$, respectively, and $L_j = T_{aq}/\tau_j$ for j $\in \{sq,i,b\}$. 

In classical SPI, an input coherent state $|\psi\rangle = |\alpha\rangle$ transforms into $|\sqrt{\eta_d\eta_0\tilde{\chi}^{(k)}}\alpha\rangle$ by an effective BS, and the photon statistics follows a Poisson distribution with mean and variance, $\bar{N}_{sc}$ = $\sigma_{sc}^2$ = $\eta_d\eta_0\tilde{\chi}^{(k)}|\alpha|^2$=$\eta_d\eta_0\tilde{\chi}^{(k)}\bar{n}_{sc}$. 
The subscript $c$ stands for the classical state. 
For CSPI and PSPI, we can similarly calculate the accumulated signal photon number $\mu_{sc}$ during the pulse width $T_w$:
\begin{equation}
\begin{split}
&\mu_{sc} \sim N(L_{sc}\bar{N}_{sc}, \:\:L_{sc}\sigma_{sc}^2),\\
\end{split}
\end{equation}
where $L_{sc} = T_w/\tau_{sc}$, and $\tau_{sc}$ is the coherence time of signal photons in classical imaging.
We set $L_{sc}\bar{N}_{sc}$ = $L_{sq}\bar{N}_{sq}$ to match the transmitter power of quantum and classical schemes equally.

The next step is to calculate the post-processed photon number for each scheme.
In QSPI, based on the coincidence measurement between photon-pairs, the accidental noise-idler coincidence counts are $\mu_b\mu_iT_{c}/T_{aq}$.    
Given the heralding efficiency $\eta_h$, the signal-idler coincidence counts are $\eta_h\mu_{sq}$. 
On the other hand, PSPI only selects the photons existing in the pulse. The total number of photons arriving at the detector within the pulse width $T_w$ is therefore given by the sum of noise photon number $\mu_bT_w/T_{aq}$ and signal photon number $\mu_{sc}$. 
Finally, if we take account of the Poissonian photoelectric process of a detector \cite{QOptics}, the final photon counts can be obtained as:
\begin{equation}\label{photoncounts}
\begin{split}
&I_c  \:\sim \text{Poi}\:(\lambda = \mu_{sc} + \mu_b),\\
&I_q  \:\sim \text{Poi}\:(\lambda = \mu_{sq}\eta_h + \mu_b\mu_iT_{c}/T_{aq} ),\\
&I_p  \:\sim \text{Poi}\:(\lambda = \mu_{sc} + \mu_bT_w/T_{aq}),
\end{split}
\end{equation}
where $\lambda$ is a parameter of Poisson distribution.
The images of CSPI, QSPI, and PSPI can be reconstructed by substituting $I_c$, $I_q$, and $I_p$ into Eq. \eqref{Cov}. 
As shown in Eq. \eqref{photoncounts}, the noise in QSPI and PSPI processes can be reduced by a factor of $\mu_iT_{c}/T_{aq}$ and $T_w/T_{aq}$, respectively, and therefore PSPI achieves a comparable noise rejection to QSPI when we take $T_w \sim \mu_iT_{c}$.

%%=========================== experiments and results==============================
\section{Experiments and results}

\begin{figure*}[h]
\centering
\includegraphics[width = 1\textwidth]{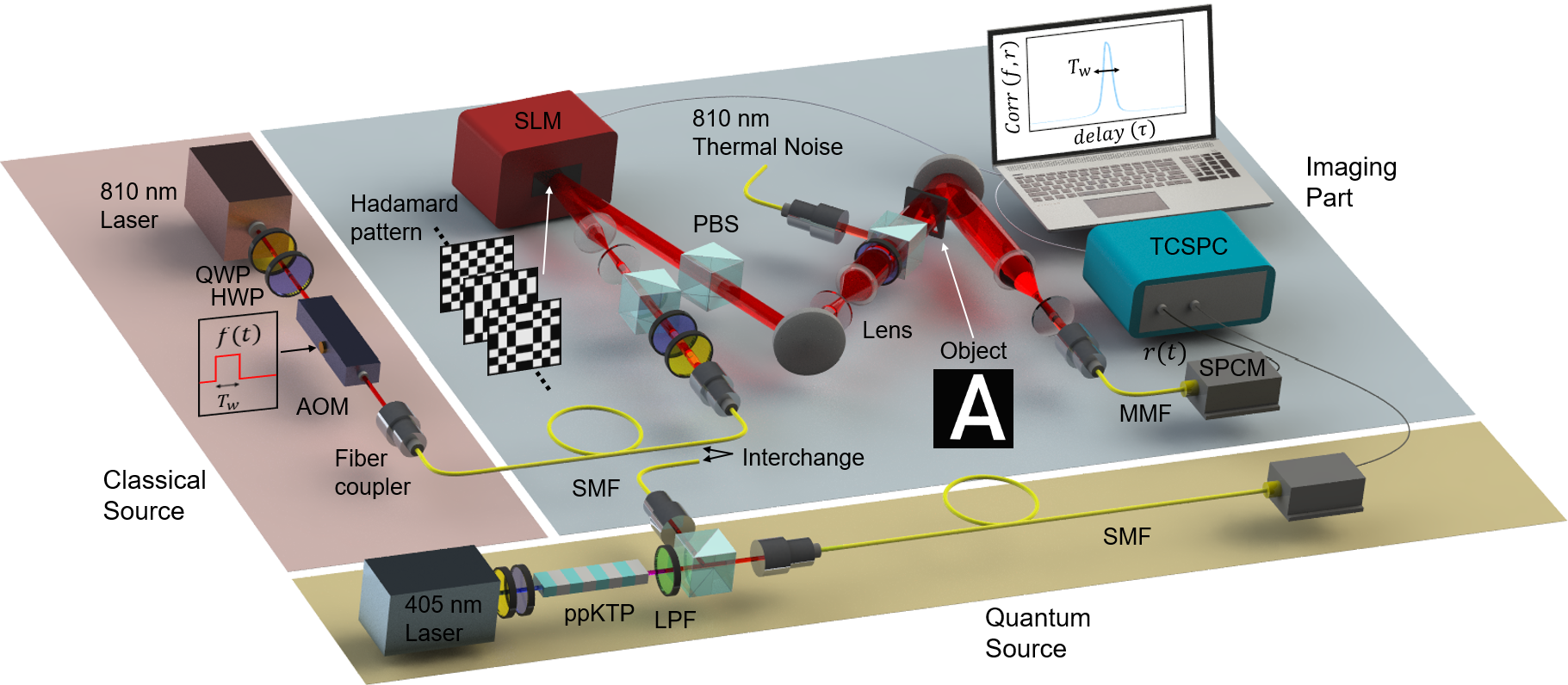}
\caption{Schematic of experimental setup. For PSPI, an 810 nm CW laser is modulated by an AOM to prepare a pulsed signal. The AOM driving pulse $f(t)$ is generated by a delay generator with pulse width $T_w$ and pulse duration $T_{aq}$. For QSPI, 810 nm degenerate down-converted photon-pairs are used. While idler photons are directly sent to an SPCM to herald signal photons in QSPI, the signal photons are projected onto the SLM and spatially modulated according to a Hadamard pattern in both classical and quantum schemes. The modulated signal photons are then sent to the target object, which is subject to an incoherent thermal background noise. After the interaction with the target object `A', both signal and noise photons are detected by an SPCM and analyzed with TCSPC. Finally, the object image is reconstructed with the modulation patterns and photon count for each pattern. LPF, long-pass filter; QWP, quarter waveplate; HWP, half waveplate; PBS, polarizing beam splitter; SMF, single-mode fiber; MMF, multi-mode fiber.}\label{imagingsetup}
\end{figure*}

The experimental schematics of CSPI, QSPI, and PSPI are shown in Fig. \ref{imagingsetup}. 
An attenuated 810 nm continuous-wave (CW) laser is used for CSPI and PSPI. 
Pulsed signal photons are generated by an acousto-optic modulator (AOM) with an electrical driving pulse signal $f(t)$, which is formed by a delay generator (Stanford Research System, INC, DG535). 
The pulse duration is set to be equal to the acquisition time $T_{aq}$, and the pulse width $T_w$ is 1 ms or 75 ms. 
The linearly polarized beam is first expanded and sent to the SLM (Thorlabs, EXULUS-HD1/M) panel of which each pixel modulates the polarization horizontally or vertically according to a Hadamard pattern.
Therefore, after the modulated beam interacts with the PBS, the spatial beam profile becomes same as the Hadamard pattern.
For each 1.5 s pulse duration, the signal photon count is 12000 when the pattern is fully bright.
The image is obtained with a 32 $\times$ 32 pixels resolution.
To completely reconstruct a 32 $\times$ 32 pixels image with a reflective SLM, one needs total 2$\times$1024 = 2048 Hadamard patterns including 1024 inversed-patterns.
This is because the reflective SLM cannot realize the negative components of Hadamard matrix, and we can only use positive weighting factors \cite{OE20Gibson}. Here, we reduced the number of modulation patterns to 2$\times$350 = 700 with compressive methods, resulting in a 3 times faster image reconstruction \cite{SciRep17Sun}.

After going through the PBS, the modulated signal photons illuminate the target object `A' with a size of 1 cm $\times$ 1 cm, which is embedded in a noisy environment. 
To simulate the background noise, we add a pseudo-thermal light generated by focusing an independent 810 nm laser beam onto a rotating ground-glass disk and combined to the signal photons through the PBS \cite{PL66Arecchi,PRL17Ihn}. 
Finally, both signal and noise photons are coupled to an MMF and detected by a single photon counting module (SPCM) (Excelitas, SPCM-NIR-13).
Each detected photon is time-tagged and counted by a time-correlated single photon counting (TCSPC) (qutools, quTAU).
The acquisition time $T_{aq}$ for each pattern is 1.5 s, and the pattern-switching time of the SLM is 1 s. 
The overall channel efficiency for the signal path $\eta_{0}$ is about 1.5\%, and the detector efficiency $\eta_d$ is 65\%.
The low channel efficiency is mainly due to the loss of signal photons at the SLM panel and imperfection of modulation. 

\begin{figure}[t]
\centering
\includegraphics[width = 0.8\textwidth]{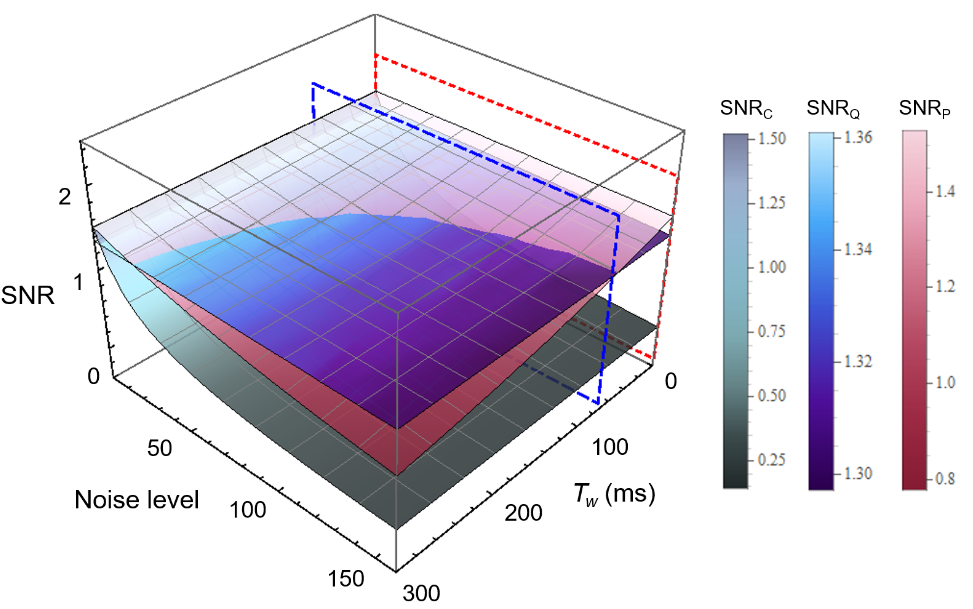}
\caption{Numerically simulated image SNRs of CSPI (green),  QSPI (blue), and PSPI (red) as a function of the thermal noise level and the pulse width. QSPI and PSPI show remarkable noise robustness compared to CSPI. 
The red dotted and blue dashed planes correspond to $T_w =$ 1 ms and 75 ms, respectively, which are experimentally demonstrated.}
\label{surfaceplot}
\end{figure}

In QSPI, a periodically poled potassium titanyl phosphate (ppKTP) crystal pumped by a 405 nm CW laser generates 810 nm collinear-degenerate photon-pairs via type-II spontaneous parametric down-conversion (SPDC).
For the SPDC source, single counts rate is 550 kcps at 9.2 mW pump power and the heralding efficiency $\eta_h$ is about 13$\%$. 
Here, heralding efficiency refers to the ratio of coincidence counts to signal counts. 
Note that the source properties are measured without the imaging setup.
The signal and idler photons are split by a PBS and coupled to SMFs. 
Signal photons are sent to the SPI setup and idler photons are directly sent to the SPCM for heralding.  
The received signal photon counts after the SPI channel are set as 12000 counts per 1.5 s for every scheme when the pattern is fully bright. 
For the PSPI, we send one pulse per 1.5 s, and signal photons exist only within the pulse width. 
Therefore, one pulse has 12000 counts on average. 
For the CSPI and QSPI, signal photons exist in all 1.5 s time duration, and the single counts of signal photons during 1.5 s are 12000 on average. 
Therefore, the averaged signal intensities used for 1.5 s acquisition were matched in three schemes.
For a timing jitter of SPCM ($\sim$300 ps) and biphoton temporal separation $\tau_{two} (<$ 1 ps), we set the coincidence window $T_{\text{c}}$ = 650 ps to ensure that all photon-pairs contribute to QSPI process \cite{Optica19Liu}.

\begin{table}[b]
\centering
\caption{Device parameters used in both theory and experiment}
\begin{tabular}{l ccccccccc}
\hline
Parameter & $T_w$ & $T_{aq}$ & $T_{c}$ & $\tau_{sc}$ & $\tau_{sq}(=\tau_{i})$ & $\tau_{b}$ & $\eta_d$ & $\eta_h$ & $\eta_0$ \\
\hline
Value & 1 ms & 1.5 s & 650 ps & 13.3 $\mu$s & 5 ps & 681 ns & 0.65 & 0.12 & 0.015 \\
& 75 ms &  &  &   && &&  &\\
\hline
\end{tabular}
 \label{parameters}
\end{table}

In order to reconstruct images, single counts ($I_c$), coincidence counts ($I_q$), and post-selected single counts ($I_p$) are measured for each imaging scheme.
In CSPI, we use the total photon counts during the acquisition time; $I_c^{(k)}$ = $\int_0^{T_{aq}} r(t)dt$, where $r(t)$ represents the detected photon counts at time $t$.
In QSPI, the coincidence count is extracted from the second-order correlation function, $g^{(2)}{(\tau)}$, between the noisy signal photons and idler photons. 
With this second-order correlation measurement, we do not have to exactly match the signal-idler path lengths equally. 
The peak of $g^{(2)}{(\tau)}$ appears at the delay point $\tau_0$, which corresponds to the path length difference between the signal and idler paths; $I_q^{(k)} = g^{(2)}(\tau_0)$.  
Finally, in PSPI, the cross-correlation function between the AOM driving pulse $f(t)$ and $r(t)$ is calculated by $C(\tau)$ = $\int_0^{T_{aq}} f(t)r(t+\tau)dt$. 
Since signal photons exist only within the pulse, we obtain noise-filtered signal counts from the maximum value of cross-correlation function; $I_{p}^{(k)} = \text{max}\{C(\tau)\}$. 
Then, the object image is retrieved by calculating $G^{(2)}(i,j)$ between the photon counts \{$I_{c}^{(k)}$\}, \{$I_{q}^{(k)}$\}, or \{$I_{p}^{(k)}$\} and modulation patterns $\{P^{(k)}(i,j)\}$. 
Table. \ref{parameters} presents device parameters used in both theory and experiment.

\begin{figure}[t]
\centering
\includegraphics[width = 0.8\textwidth]{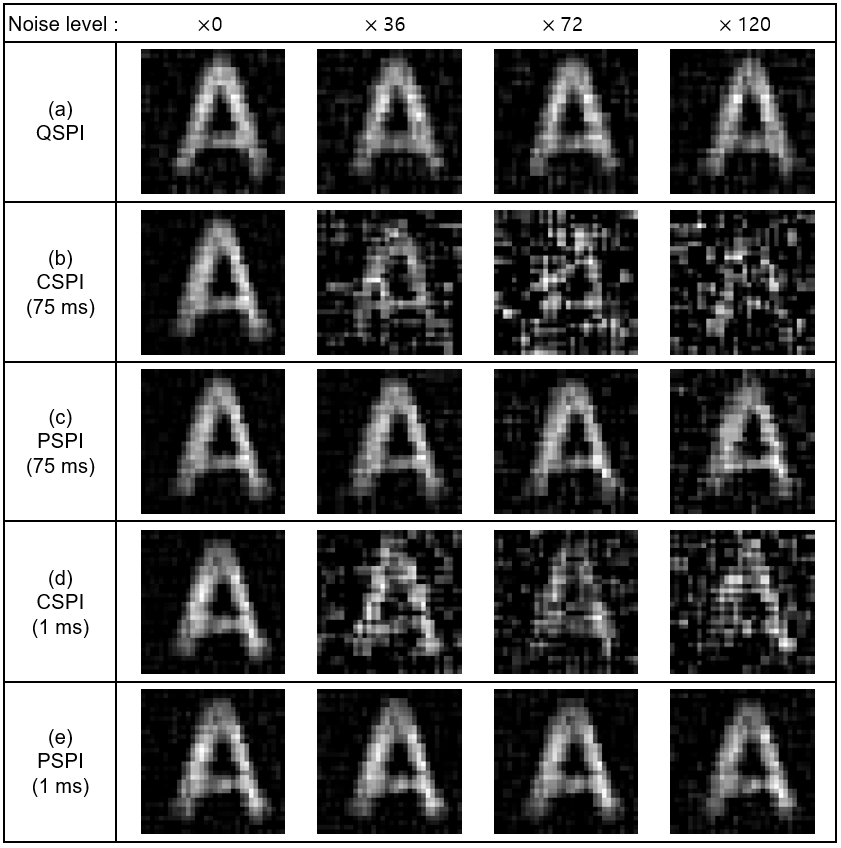}
\caption{Experimental images obtained by (a) QSPI, (b) CSPI and (c) PSPI with $T_w$ = 75 ms, (d) CSPI and (e) PSPI with $T_w$ = 1 ms in different noise levels. The noise level means the ratio of average noise photon counts to average signal photon counts. All images are reconstructed with 700 Hadamard patterns.} 
\label{images}
\end{figure}

We quantify the image quality by using a signal-to-noise ratio (SNR) \cite{SciAdv20Barzanjeh}:
\begin{equation}\label{SNR}
\text{SNR} = \frac{|\:\mu_{\text{O}} - \mu_{\text{B}}\:|^2}{2\:(\sigma_{\text{O}} + \sigma_{\text{B}})^2}
\end{equation}
where $\mu_{i}$ and $\sigma_{i}$ ($i = \text{O, B}$) are the mean and standard deviation of gray scale pixel values comprising the object region (O) and background region (B) in the image, respectively. 
Fig. \ref{surfaceplot} shows numerically calculated SNRs of CSPI, QSPI and PSPI by using device parameters in Table \ref{parameters}.
Noise-robustness of each imaging scheme is compared with respect to background thermal noise level and the pulse width. 
The noise level is represented by the ratio of average noise photon counts rate and signal photon counts rate.
When $T_w$ is less than 80 ms, PSPI shows a higher SNR than that of QSPI at the noise level less than 100. 
This is mainly due to the low heralding efficiency of quantum source.
In this work, two cases of $T_w=$ 1 ms and 75 ms, marked with red dotted and blue dashed planes in Fig. \ref{surfaceplot}, are experimentally demonstrated.

Fig. \ref{images} shows the reconstructed images of CSPI, QSPI and PSPI at the noise level from 0 to 120. 
As the background noise level increases above 36, CSPI presents severely deteriorated images (see Fig.\ref{images}(b) and (d)). 
On the other hand, QSPI and PSPI provide clear object images for all noise levels as shown in Fig. \ref{images}(a), (c), and (e). 
It shows that both QSPI and PSPI based on temporal correlation measurements are highly effective in noise suppression.
We also note that the short pulse width ($T_w=$ 1 ms) leads to higher background noise rejection than the long pulse case with $T_w$ = 75 ms as shown in Fig. \ref{images}(c) and (e).

\begin{figure}[t]
\centering
\includegraphics[width = 1\textwidth]{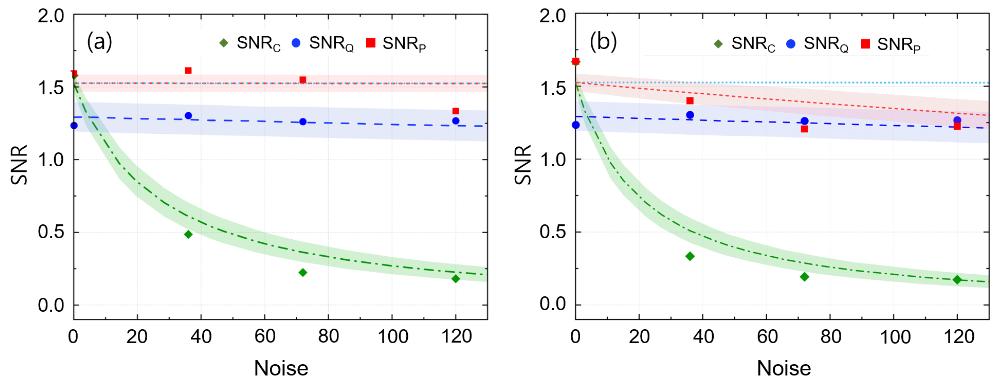}
\caption{SNRs of QSPI, CSPI, and PSPI images with pulse widths (a) $T_w$ = 1 ms and (b) $T_w$ = 75 ms. Note that the SNRs of QSPI plotted in (a) and (b) are identical because QSPI is independent of $T_w$. Solid markers represent experimental data, and lines and shaded regions describe theoretical values from numerical simulations. Light blue dotted line is the theoretical average SNR of QSPI when $\eta_h=1$.}
\label{SNR}
\end{figure}

Fig. \ref{SNR} shows numerically calculated and experimental SNRs of images reconstructed by CSPI (SNR$_C$, green diamond), QSPI (SNR$_Q$, blue circle), and PSPI(SNR$_P$, red square).
Fig. \ref{SNR}(a) and (b) correspond to $T_w$ = 1 ms and 75 ms cases, respectively, which are cross sectional planes marked in Fig. \ref{surfaceplot}.
The shaded area denotes an uncertainty, which is 10 to 90 percentile interval of 5000 numerical SNR values, and the center lines in shaded regions refer to average values of numerical simulations.
The slight discrepancies between theoretical curves and experimental values of CSPI and PSPI are due to the slow drift of the laser power during the imaging process.
We experimentally observe the classical SNRs that surpass quantum-enhanced SNRs in the presence of noise, which is predicted in Fig. \ref{surfaceplot}.
As mentioned earlier, the heralding efficiency less than 1 is the main reason of the SNR degradation of QSPI.
In Fig. \ref{SNR}, light blue dotted lines describe the SNRs of QSPI when the heralding efficiency is $\eta_h=1$.
Especially for 1 ms pulse width, we can obtain a nearly unchanged SNR of PSPI even in the presence of 100 times larger background noise as shown in Fig. \ref{SNR}(a).

\section{Discussion}

From the results above, one may wonder if a classical imaging with a proper pulse width can always give SNR enhancement that is comparable to, or even better than quantum imaging.
It is worth pointing out that this can be possible under proper assumptions. 
In this work, we assumed a stationary noise system where the photon statistics of thermal background noise does not change over the acquisition time. 
Under this assumption, the accidental noise count in PSPI is approximately $\mu_bT_w/T_{aq}$ for every time interval of $T_w$, which gives a distinctive peak of the cross-correlation function.
However, if noise has an arbitrary pulsed shape, or jamming pulses exist, unwanted correlations with the signal pulses are likely to occur, which lowers the SNRs of images.
On the other hand, QSPI offers noise-robust images against both arbitrary pulsed noise and jamming pulses because signal photons probabilistically generated from a quantum source are nearly independent of any type of noise. Furthermore, it has been proved that optimized quantum receivers enable quantum illumination to surpass the fundamental limit of classical approaches in terms of target sensitivity \cite{PRA09Guha,PRL17Zhuang, PRR21Jo}. This implies we can further improve the noise-robustness of quantum imaging. 
However, despite these promising potentials of quantum imaging, it is still challenging to make a highly bright and efficient quantum source. 
Therefore, for a stationary noise environment, our work provides a classical imaging method that is easily implementable without nonlinear elements and achieves the desired noise-robustness when using quantum imaging.

\section{Conclusion}
In conclusion, we have experimentally demonstrated a classically enhanced single-pixel imaging in photon-counting regime by applying the noise suppression concept of quantum imaging. 
Here, we show remarkably enhanced performances of classical imaging at heavy noise conditions and compare the noise-robustness of CSPI, QSPI, and PSPI in terms of SNR. 
A proper selection of pulse width enables PSPI to achieve a comparable noise suppression to QSPI under a stationary noise environment. 
This work provides an efficient and simple way for noise-robust imaging in real-world scenarios.

\section*{Disclosures}
The authors declare no conflicts of interest.


\begin{thebibliography}{00}


\bibitem{NatPho10Brida} G. Brida, M. Genovese, and R. Berchera, "Experimental realization of sub-shot-noise quantum imaging," {\it Nat. Photonics} {\bf 4}, 227 (2010).
\bibitem{APL15Xu} D.-Q. Xu, X.-B. Song, H.-G. Li, D.-J. Zhang, H.-B. Wang, J. Xiong, K. Wang, "Experimental Observation of sub-Rayleigh quantum imaging with a two-photon entangled source,"{\it Appl. Phys. Lett.} {\bf 106}, 171104 (2015).

\bibitem{Science08Lloyd} S. Lloyd, "Enhanced sensitivity of photodetection via quantum illumination," \textit{Science} {\bf 321}, 1463-1465 (2008).
\bibitem{PRL08Tan} S.-H. Tan, B. I. Erkmen, V. Giovannetti, S. Guha, S. Lloyd, L. Maccone, S. Pirandola, and J. H. Shapiro, "Quantum illumination with gaussian states," {\it Phys. Rev. Lett.} {\bf 101}, 253601 (2008).
\bibitem{NP10Kacprowicz} M. Kacprowicz, R. Demkowicz-Dobrzanski, W. Wasilewski, K. Banaszek, and I. A. Walmsley, "Experimental quantum-enhanced estimation of a lossy phase shift," {\it Nat. Photonics} {\bf 4}, 357 (2010).
\bibitem{PRL13Lopaeva} E. D. Lopaeva, I. Ruo Berchera, I. P. Degiovanni, S. Olivares, G. Brida, and M. Genovese, "Experimental realisation of quantum illumination," {\it Phys. Rev. Lett.} {\bf 110}, 153603 (2013).
\bibitem{PRL15Zhang} Z. Zhang, S. Mouradian, F. Wong, and J. Shapiro, "Entanglement-Enhanced Sensing in a Lossy and Noisy Environment," {\it Phys. Rev. Lett.} {\bf 114}, 110506 (2015).
\bibitem{PRA21Lee} S.-Y. Lee, Y. S. Ihn, and Z. Kim, "Quantum illumination via quantum-enhanced sensing," {\it Phys. Rev. A} {\bf 103}, 012411 (2021).

\bibitem{SciAdv19Defienne} H. Defienne, M. Reichert, J. W. Fleischer, and D. Faccio, "Quantum image distillation," {\it Sci. Adv.} {\bf 5}, eaax0307 (2019).
\bibitem{SciAdv20Gregory} T. Gregory, P.-A. Moreau, E. Toninelli, and M. J. Padgett, "Imaging through noise with quantum illumination," {\it Sci. Adv.} {\bf 6}, eaay2652 (2020).


\bibitem{PRA08Shapiro} J. H. Shapiro, "Computational ghost imaging," {\it Phys. Rev. A} {\bf 78}, 061802 (2008).
\bibitem{NP19Edgar} M. P. Edgar, G. M. Gibson, and M. J. Padgett, "Principles and prospects for single-pixel imaging," {\it Nat. Photonics} {\bf 13}, 13 (2019).

\bibitem{PRA19England} D. G. England, B. Balaji, and B. J. Sussman, "Quantum-enhanced standoff detection using correlated photon pairs," {\it Phys. Rev. A} {\bf 99}, 023828 (2019).
\bibitem{APL20Yang} J.-Z. Yang, M.-F. Li, X.-X. Chen, W.-K. Yu, and A.-N. Zhang, "Single-photon quantum imaging via single-photon illumination," {\it Appl. Phys. Lett.} {\bf 117}, 214001 (2020).
\bibitem{21Kim} J. Kim, T. Jeong, S.-Y. Lee, D. Y. Kim, D. Kim, S. Lee, Y. S Ihn, Z. Kim, and Y. Jo, "Heralded single-pixel imaging with high loss-resistance and noise-robustness," (2021) submitted.

\bibitem{SciRep18Liu} X. Liu, J. Shi, X. Wu, and G. Zeng, "Fast first-photon ghost imaging," {\it Sci. Rep} {\bf 8}, 5012 (2020).
\bibitem{AppOpt15Yang} Y. Yang, J. Shi, J. Peng, and G. Zeng, "Computational imaging based on time-correlated single-photon-counting technique at low light level," {\it Appl. Opt} {\bf 54}, 9277 (2015).
\bibitem{OE20Liu} X. Liu, J. Shi, L. Sun, Y. Li, and J. Fan, "Photon-limited single-pixel imaging," {\it Opt. Express} {\bf 28}, 8132 (2020).


\bibitem{OE20Gibson} G. M. Gibson, S. D. Johnson, and M. J. Padgett, "Single-pixel imaging 12 years on: a review," {\it Opt. Express} {\bf 28}, 28190 (2020).
\bibitem{SciRep17Sun} M.-J. Sun, L.-T. Meng, M. P. Edgar, M. J. Padgett, and N. Radwell, "Russian dolls ordering of the Hadamard basis for compressive single-pixel imaging," {\it Sci. Rep.} {\bf 7}, 3464 (2017).


\bibitem{PL66Arecchi} F. T. Arecchi, E. Gatti, and D. D. Sona, "Time distribution of photons from coherent and Gaussian sources," {\it Phys. Lett.} {\bf 20}, 27 (1966).
\bibitem{PRL17Ihn} Y. S. Ihn, Y. Kim, V. Tamma, and Y.-H. Kim, "Second-order temporal interference with thermal light: Interference beyond the coherence time," {\it Phys. Rev. Lett.} {\bf 119}, 263603 (2017).

\bibitem{Optica19Liu} H. Liu, D. Giovannini, H. He, D. England, B. J. Sussman, B. Balaji, and A. S. Helmy, "Enhancing LIDAR performance metrics using continuous-wave photon-pair sources," {\it Optica} {\bf 6}, 1349 (2019).

\bibitem{Optica17Rafsanjani} 
S. M. H. Rafsanjani, M. Mirhosseini, O. S. Magna$\tilde{\text{N}}$A-Loaiza, B.T. Gard, R. Birrittella, B. E. Koltenbah, C. G. Parazzoli, B. A. Capron, C. C. Gerry, J. P. Dowling, and R. W. Boyd et al. Quantum-enhanced interferometry with weak thermal light. \textit{Optica} {\bf 4}, 487-491 (2017)

\bibitem{QOptics} R. Loudon, {\it The Quantum Theory of Light\/} (Oxford Sci. Publications, Oxford, 2000).

\bibitem{SciAdv20Barzanjeh} S. Barzanjeh, S. Pirandola, D. Vitali, and J. M. Fink, "Microwave quantum illumination using a digital receiver," {\it Sci. Adv.} {\bf 6}, eabb0451 (2020).


\bibitem{PRR21Jo} Y. Jo, S. Lee, Y. S. Ihn, Z. Kim, and S.-Y. Lee, "Quantum illumination receiver using double homodyne detection," {\it Phys. Rev. Research.} {\bf 3}, 013006 (2021).
\bibitem{PRA09Guha} S. Guha, and B. I. Erkmen, "Gaussian-state quantum-illumination receivers for target detection," {\it Phys. Rev. A} {\bf 80}, 052310 (2009).
\bibitem{PRL17Zhuang} Q. Zhuang, Z. Zhang, and J. H. Shapiro "Optimum mixed-state discrimination for noisy entanglement-enhanced sensing," {\it Phys. Rev. Lett} {\bf 118}, 040801 (2017).


\end{thebibliography}
\end{document}